# scientific data

**OPEN**

**DATA DESCRIPTOR**

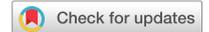

## ZTBus: A Large Dataset of Time-Resolved City Bus Driving Missions

Fabio Widmer[1,2] ✉, Andreas Ritter[1,2] & Christopher H. Onder[1]

This paper presents the Zurich Transit Bus (ZTBus) dataset, which consists of data recorded during driving missions of electric city buses in Zurich, Switzerland. The data was collected over several years on two trolley buses as part of multiple research projects. It involves more than a thousand missions across all seasons, each mission usually covering a full day of operation. The ZTBus dataset contains detailed information on the vehicle's power demand, propulsion system, odometry, global position, ambient temperature, door openings, number of passengers, dispatch patterns within the public transportation network, etc. All signals are synchronized in time and include an absolute timestamp in tabular form. The dataset can be used as a foundation for a variety of studies and analyses. For example, the data can serve as a basis for simulations to estimate the performance of different public transit vehicle types, or to evaluate and optimize control strategies of hybrid electric vehicles. Furthermore, numerous influencing factors on vehicle operation, such as traffic, passenger volume, etc., can be analyzed in detail.

## Background & Summary

Public transportation is an effective solution for reducing traffic in growing cities. It significantly reduces the number of vehicles on the road, resulting in less congestion, shorter travel times, minimal ecological footprint, and reduced overall energy consumption. In the near future, the need for such efficient urban transportation systems is likely to increase, as an estimated two-thirds of the world's population is expected to live in cities by 2050[1].

In this context, detailed driving and operational data are of great value to assist cities and transportation operators in making informed decisions on the vehicles' ideal propulsion technology and charging strategy for the respective public transportation network. Furthermore, for the development and tuning of intelligent vehicle state estimation algorithms or energy management strategies, time-resolved data of the traction system is necessary for both the vehicle manufacturers and the research community. While there are publicly available datasets describing urban traffic conditions and human mobility[2–4], as well as time-series data of personal cars[5–7] or taxis[8], publicly available time-series data of urban transit buses is lacking.

The goal of this publication is to fill this gap by presenting the ZTBus dataset[9], which is composed of data recorded throughout the course of the projects «SwissTrolley plus»[10] and ISOTHERM[11], both of which were collaborations between industry partners and public research institutions and were financially supported by the Swiss Federal Office of Energy (SFOE). The dataset covers more than a thousand driving missions of two trolley buses that were in operation between April 2019 and December 2022. It consists of detailed time series that represent the power demand, propulsion system, odometry, global position, ambient temperature, door openings, number of passengers, and the dispatch patterns within the public transportation network of the two vehicles. The time series are provided in a synchronized form and are sampled every second. Aggregated quantities for each of the missions are provided in a metadata table. A schematic overview of the data acquisition and curation procedure, which is explained in greater detail below, is shown in Fig. 1. Figure 2 presents the full extent of the dataset.

This data offers the potential to be used in a broad variety of fields. For example, the time-resolved global navigation satellite system (GNSS) data can be combined with odometry signals, such as the wheel speeds and the steering angle, and processed using sensor fusion approaches. Such algorithms can significantly improve the raw pose estimates provided by the GNSS sensor, and facilitate the use of dead reckoning approaches in case of momentary signal outage. Additionally, the large amount of data on a set of given routes is suitable for the examination of algorithms for trajectory filtering and map matching in machine learning contexts.

[1]ETH Zurich, Institute for Dynamic Systems and Control, Zurich, 8092, Switzerland. [2]These authors contributed equally: Fabio Widmer, Andreas Ritter. ✉e-mail: fawidmer@idsc.mavt.ethz.ch

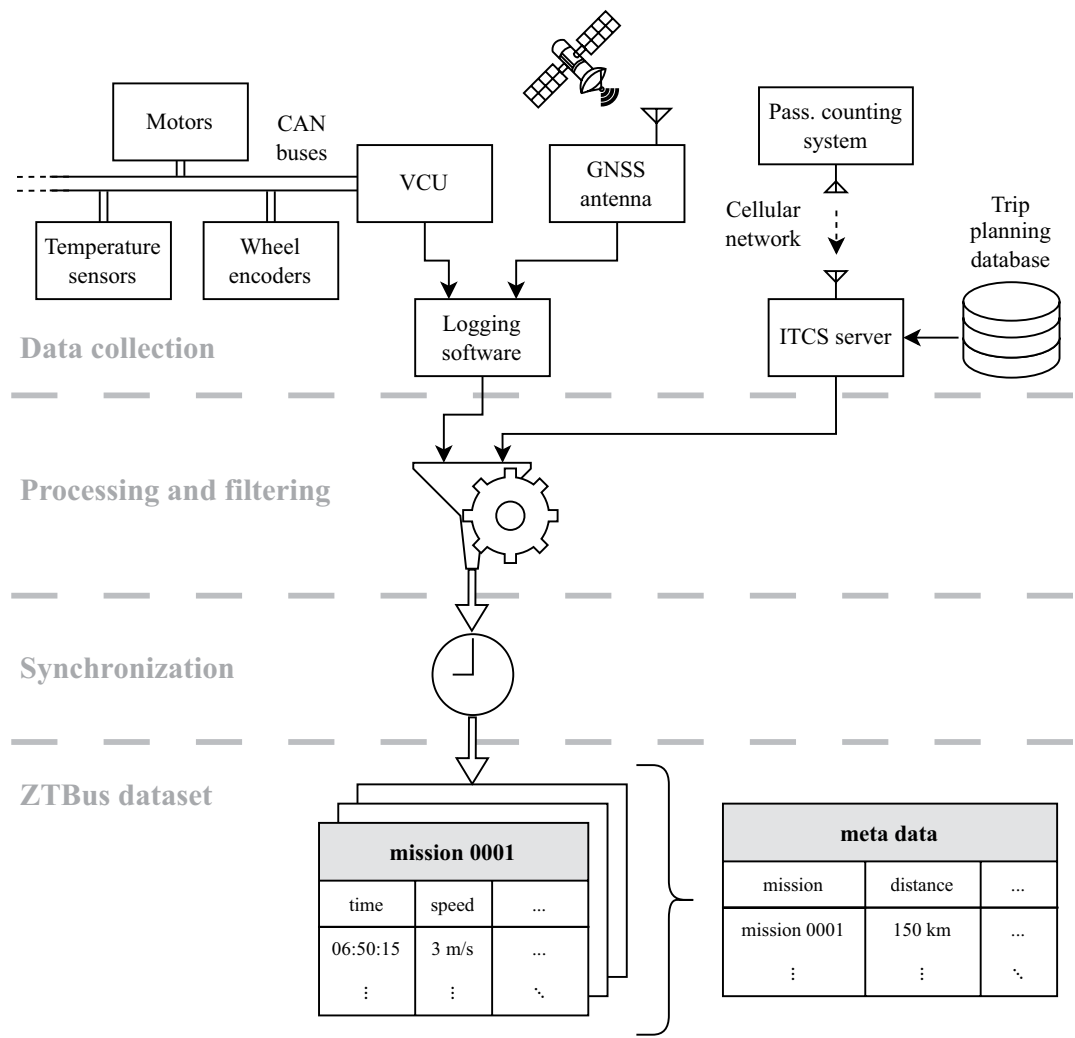

**Fig. 1** Data acquisition and curation. Signals from three different sources, i.e., the vehicle control unit (VCU), the global navigation satellite system (GNSS) antenna, and the passenger counting system, are used in the definition of the driving missions. Various filtering steps are added to reject erroneous and unrepresentative data. Finally, time synchronization and sampling are performed to present the data in a tabular format.

Machine learning may also be utilized to predict various influence factors in public transportation, such as the number of passengers that travel a certain distance at a given time, the traffic levels on specific roads and at specific times of the day, or the expected speed profiles of the vehicles in the near future or in general on certain road segments.

Finally, the aggregated data enables the examination of long-term correlations such as the impact of COVID-19 mitigation measures on passenger numbers, the effect of weather conditions on energy consumption, etc.

The dataset presented in this manuscript has been used in several of our own publications in the context of the research activities mentioned above: (1) The position, odometry, and velocity data served to develop and evaluate a real-time incremental graph construction algorithm[12]. (2) Time-resolved speed, torque, and braking pressure signals were used for the development of the model-based vehicle mass and road grade estimation method[13]. (3) The spatio-temporal nature of the power request signal was used to quantify the relation between grid and battery energy usage on certain road segments, which then served to derive a stochastic model predictive control approach[14]. (4) The optimal design and control of a thermal energy buffer in an electric city bus was studied based on the passenger loads, velocity and elevation profiles[15]. (5) A set of 16 representative all-day driving missions served to optimize the bus operation in terms of managing battery degradation throughout the vehicle lifetime[16]. (6) Hourly-averaged data was used to conduct a large-scale sensitivity analysis of the thermal comfort systems, allowing a comparison of various heating, ventilation and air conditioning (HVAC) systems[17].

## Methods

**Data collection.** The ZTBus dataset[9] was recorded on two trolley buses during regular operation by Verkehrsbetriebe Zurich (VBZ). Both are "HESS lighTram® 19 DC" buses, which are single-articulated, have an overall length of about 19 m, a curb weight of about 19 t, and a maximum passenger capacity of about 160. They are equipped with traction batteries, which allow them to run for a few kilometers without the overhead power grid. The dataset covers the operation of the buses on various bus routes in Zurich's public transportation network, which are visualized in Fig. 3.

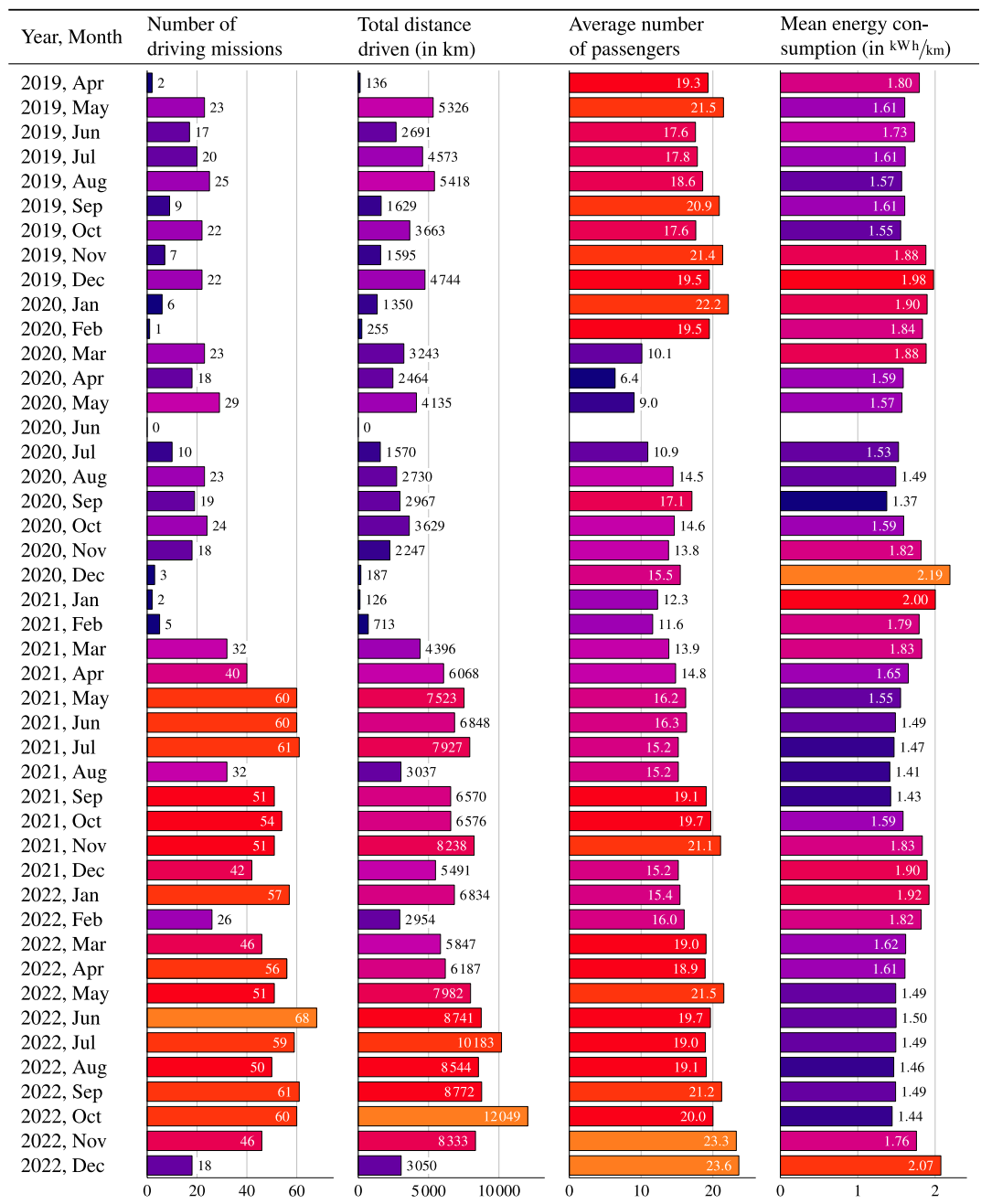

**Fig. 2** Visualization of the extent of the ZTBus dataset[9], which includes a total of 1409 driving missions over the period of over 3.5 years between April 2019 and December 2022.

The data included in the ZTBus dataset[9] originates from the three systems described below and schematically shown in Fig. 1. It is recorded via onboard logging systems specifically developed for that purpose.

1. The majority of the data is provided by the vehicle control unit (VCU) to which the raw measurement data is directly transmitted via multiple controller area network (CAN) buses from the various vehicle components. As this data is used during the normal operation of the bus, these signals are always available if the attached logging system works as intended.
2. The data related to the global positioning of the vehicles is provided by a GNSS antenna mounted on their roofs. The GNSS data may be temporarily unavailable if no reliable connection to the satellites can be established, which may be the case, e.g., during bad weather, between tall buildings, or in underpasses.
3. The passenger counts are estimated via onboard infrared-based passenger counting systems that transmit their estimates to the public transportation operator's server computer via the local cellular network. This data is then automatically synchronized and augmented with the data from the intermodal transport control system (ITCS), i.e., the corresponding bus route number and stop names. We refer to this combined data as "ITCS data".

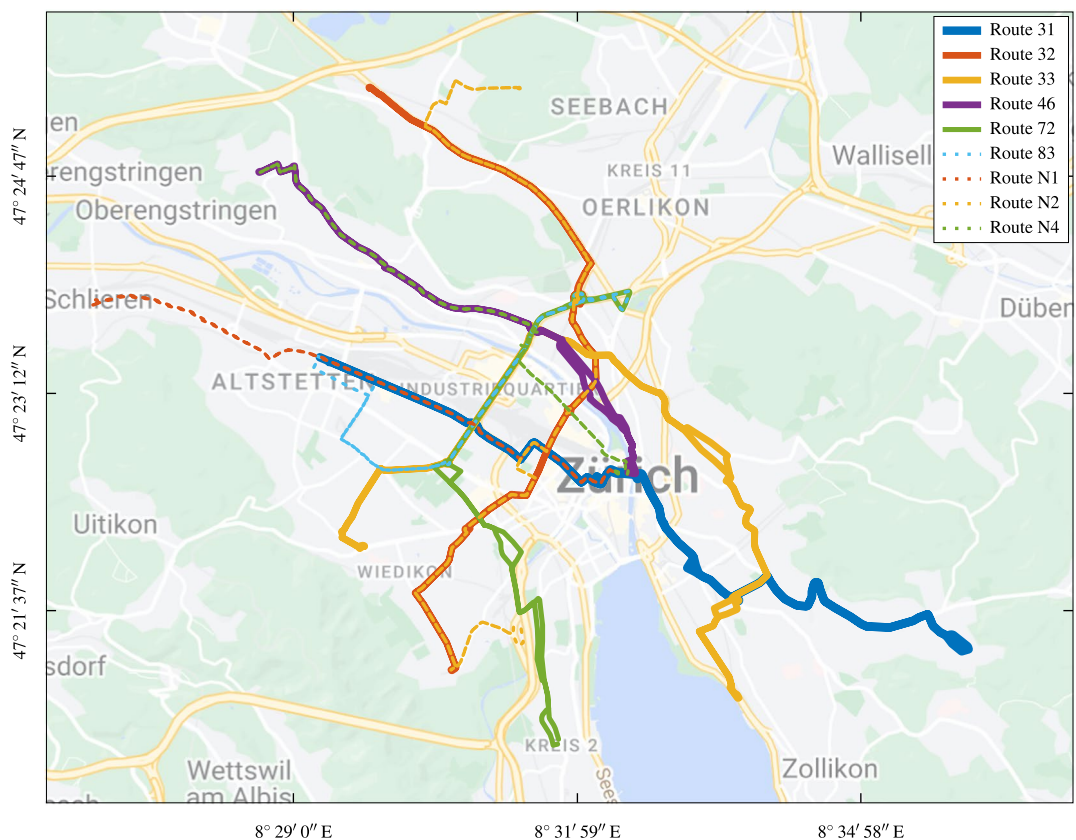

**Fig. 3** Path of the nine bus routes covered by the dataset. The route data is publicly available[18]. Map data ©2023 Google.

The data is organized in "driving missions", which we define as the entire period from the moment the bus is switched on until the moment it is switched off.

**Selection of data records.** To ensure that the dataset meets the requisite integrity and quality standards, we include only those records that represent complete driving missions during regular public transportation operation. For example, we reject test drives, short trips within depots, and missions that are completely missing any data from the three systems described above. The details of this selection step are described in the section on technical validation below.

**Processing.** We aim to reduce the processing of data to a minimum. In particular, instead of applying sophisticated filtering and smoothing techniques, we publish the raw measurement data received from the sensory devices, which allows its use for the development or tuning of such smoothing and filtering algorithms as well. The processing steps that were nevertheless considered necessary and were carried out are as follows:

- On our vehicles, the most accurate indicator of the vehicle speed is provided by the rotational speed sensors mounted on the motor shafts. As we aim to present our data in a manner that is independent of the vehicle's specific drivetrain, we use an estimate of the "compound" transmission ratio $\gamma$ to convert the rotational speed measurements $\omega$ to the longitudinal vehicle speed $v$:

$$v = \frac{\omega}{\gamma}. \tag{1}$$

- The compound transmission ratio thus combines the effects of the transmission, final drive, and wheel radius. For estimating $\gamma$, we analyze measurement data of perfectly straight driving sections, where the traveled distance according to filtered GNSS data is compared to the total angle covered by the electric machine. Note that we use the rotational speed measurements of the motor at the middle axle for the calculation of the travel speed in Eq. 1. That way, errors induced due to offtracking during tight turns are minimized. The estimated transmission ratio is also used to calculate the traction force

$$F_{\text{trac}} = \gamma \cdot T_{\text{trac}}, \tag{2}$$

- where $T_{\text{trac}}$ represents the total torque provided by the electric machines. Since the vehicle acceleration in city buses is generally low so as to allow for a comfortable ride even for standing passengers, our experience shows that the error induced by tire slip is typically negligible.
- To focus on the valuable information recorded between the initial departure and the final arrival of each driving mission, we discard any data recorded more than 1 min before and more than 1 min after the actual driving.
- When combining numbers of passengers (estimated by the on-board infrared-based counting systems) with bus route and stop names (provided by the ITCS), some inconsistencies are filtered out by the public transportation operator. For example, trips are discarded if the total number of passengers boarding differs too much from the total number of passengers alighting throughout one trip, or if a recorded trip cannot be uniquely matched to a trip in the ITCS database. To exclude any remaining erroneous ITCS data, which may occur, for instance, when stops are skipped, we use the locations of the reported stops according to publicly available general transit feed specification (GTFS) data[18] and compare them to the location estimates provided by the GNSS sensors. If the locations are farther than 100 m apart for at least three stops in a row, the ITCS data associated with these stops is discarded.
- Finally, the data from the three different sources introduced above, i.e., the VCU, GNSS, and the ITCS, is synchronized and resampled. For this purpose, we first generate a new date-time vector in coordinated universal time (UTC) with a uniform sampling period of 1 s covering the time window highlighted above. The signals are then mapped to this time vector as follows:

  - The ITCS data is only given at discrete time events approximately matching the moments the bus is leaving a stop. As the processing of the raw data is to be kept to a minimum, the ITCS data is not interpolated. Instead, the nearest sample times of the new date-time vector are identified and the discrete values are mapped accordingly.
  - All binary (status) signals are interpreted as piecewise constant signals and are thus resampled via previous neighbor interpolation.
  - All other signals are linearly interpolated.

## Data Records

The ZTBus dataset is published on the repository for publications and research data of ETH Zurich[9]. It is organized in two different types of comma-separated values (CSV) text files, the first of which describes the 1409 individual driving missions, while the second contains metadata of all driving missions.

The names of the files that describe the individual driving missions are based on the vehicle identification number (either 183 or 208) and the time period in which the data was collected. For example, the data collected on the bus numbered 183 between 16 Oct 2019 02:52:43 and 16 Oct 2019 07:10:12, both given in UTC, is available in the following file:

```
B183_2019-10-16_02-52-43_2019-10-16_07-10-12.csv
```

The metadata describing all driving missions is provided as metaData.csv.

**Detailed description of the time-resolved measurement data.** The ZTBus dataset[9] consists of 1409 driving missions, each of which is described in a separate CSV file. All files have the same structure and format, where the first row contains the headers of the corresponding columns and the remaining rows describe the set of data samples recorded at a specific moment in time. This time index is represented in the first column as absolute UTC time, expressed according to ISO 8601.

The columns are described in Table 1, where NaN represents unavailable data, unless specified otherwise.

**Detailed description of the metadata.** The metadata of the driving missions is tabulated as described in Table 2. The first row contains the headers of the corresponding columns. The remaining rows contain metadata of the driving missions, indexed via the corresponding file name in the first column.

## Technical Validation

In this section, we explain the various measures we have taken to ensure the requisite integrity and quality of the ZTBus dataset[9]. In particular, we have developed the following selection criteria.

- We start by considering only records without any known issues in the logging toolchain, i.e., issues such as erroneous timestamps, or bugs in any of the involved software components. For this initial selection, we additionally reject records with corrupt file contents. Furthermore, we only consider records that span a time interval of at least 1 h each, as all shorter records do not represent any regular public transportation operation. This initial selection contains a total of 2046 missions.
- We reject missions where either VCU, GNSS, or ITCS data is unavailable throughout the entire mission, which reduces the dataset by 189 missions.
- If we detect a gap of at least 10 s in any of the VCU data, we reject the data of the entire mission, as this hints at a potential issue either in the logging toolchain or with the onboard clock. Such a gap is detected in 13 missions.

| Column name | Data type | Unit | Description |
| --- | --- | --- | --- |
| `time_iso` | datetime |  | The absolute UTC time, expressed according to ISO 8601. |
| `time_unix` | integer | s | The Unix timestamp, i.e., the number of seconds passed since 00:00:00 UTC on 1 Jan 1970. |
| `electric_powerDemand` | float | W | The overall electric power demand of the vehicle. The values include the power demand of the traction motors (which can be negative during recuperation phases) and all auxiliary power consumers like HVAC, air compressor, lighting, infotainment systems, etc. |
| `gnss_altitude` | float | m | The altitude above sea level measured by the GNSS sensor. |
| `gnss_course` | float | rad | The course (heading) provided by the GNSS sensor. Traveling north is represented by a course of 0 rad or $2\pi$ rad. Traveling east is represented by a value of $\pi/2$. When the bus is not moving, this estimate is held constant by the GNSS sensor on bus 183 and set to zero on bus 208. |
| `gnss_latitude` | float | rad | The latitude on WGS 84 measured by the GNSS sensor. |
| `gnss_longitude` | float | rad | The longitude on WGS 84 measured by the GNSS sensor. |
| `itcs_busRoute` | string |  | The VBZ bus route name, provided by the ITCS. For unavailable data, we use a dash ("–"). |
| `itcs_numberOfPassengers` | float | — | The estimated number of passengers on the bus, measured with infrared-based passenger counting systems. |
| `itcs_stopName` | string |  | The name of the bus stop served, provided by the ITCS. For unavailable data, we use a dash ("–"). |
| `odometry_articulationAngle` | float | rad | The angle of the pivoting joint (articulation). A positive angle is observed in a right turn. |
| `odometry_steeringAngle` | float | rad | The angle of the front wheels relative to the vehicle body. A positive angle is observed in a left turn. |
| `odometry_vehicleSpeed` | float | m/s | The vehicle speed. This value is based on the rotational velocity of the drive shaft of the middle axis and a compound transmission ratio according to eq. 1. |
| `odometry_wheelSpeed_fl` | float | m/s | The wheel speed of the front left wheel, measured using a wheel encoder. Due to the nature of the measurement, it is not reliable below approximately 1.5 m/s. |
| `odometry_wheelSpeed_fr` | float | m/s | The wheel speed of the front right wheel, measured using a wheel encoder. Due to the nature of the measurement, it is not reliable below approximately 1.5 m/s. |
| `odometry_wheelSpeed_ml` | float | m/s | The wheel speed of the middle left wheel, measured using a wheel encoder. Due to the nature of the measurement, it is not reliable below approximately 1.5 m/s. |
| `odometry_wheelSpeed_mr` | float | m/s | The wheel speed of the middle right wheel, measured using a wheel encoder. Due to the nature of the measurement, it is not reliable below approximately 1.5 m/s. |
| `odometry_wheelSpeed_rl` | float | m/s | The wheel speed of the rear left wheel, measured using a wheel encoder. Due to the nature of the measurement, it is not reliable below approximately 1.5 m/s. |
| `odometry_wheelSpeed_rr` | float | m/s | The wheel speed of the rear right wheel, measured using a wheel encoder. Due to the nature of the measurement, it is not reliable below approximately 1.5 m/s. |
| `status_doorIsOpen` | boolean |  | A binary flag indicating whether at least one door is open. |
| `status_gridIsAvailable` | boolean |  | A binary flag indicating whether the current collector of the trolley bus is connected to the overhead grid. |
| `status_haltBrakeIsActive` | boolean |  | A binary flag indicating whether the halt brake is active. This brake is automatically activated whenever the bus stands still. |
| `status_parkBrakeIsActive` | boolean |  | A binary flag indicating whether the park brake is active. This fail-safe brake works also if the vehicle is not powered. It can be manually activated by the driver, e.g., for longer standstill periods. |
| `temperature_ambient` | float | K | The measured ambient temperature. The temperature sensor is located behind the front bumper. Due to the sensor resolution, it is only available in increments of 1 K. |
| `traction_brakePressure` | float | pa | The mean pressure in the friction braking lines. |
| `traction_tractionForce` | float | N | An estimate of the overall traction force provided by the two traction motors. These values are based on an estimate of the traction torque provided by the two traction motors and a compound transmission ratio according to eq 2. |

**Table 1.** Column descriptions of the driving mission tables. The unit column is left blank for non-numeric data types.

| Column name | Data type | Unit | Description |
|---|---|---|---|
| name | string | | The name of the data file represented by the corresponding row of the metadata table. |
| busNumber | integer | — | The number of the bus on which the data was recorded (183 or 208). |
| startTime_iso | datetime | | The absolute UTC time of the start of the driving mission, expressed according to ISO 8601. |
| startTime_unix | integer | s | The Unix timestamp of the start of the driving mission, in seconds since 00:00:00 UTC on 1 Jan 1970. |
| endTime_iso | datetime | | The absolute UTC time of the end of the driving mission, expressed according to ISO 8601. |
| endTime_unix | integer | s | The Unix timestamp of the end of the driving mission, in seconds since 00:00:00 UTC on 1 Jan 1970. |
| drivenDistance | float | m | The distance covered during the entire driving mission. This value is calculated via trapezoidal integration of the time series odometry_vehicleSpeed. |
| busRoute | string | | The most frequently occurring value of the time series itcs_busRoute. |
| energyConsumption | float | J | The overall electric energy consumption of the vehicle during the entire driving mission. This value is calculated via trapezoidal integration of the time series electric_powerDemand. |
| itcs_numberOfPassengers_mean | float | — | The average number of passengers in the bus, i.e., the mean value of the time series itcs_numberOfPassengers, where NaN values are omitted. |
| itcs_numberOfPassengers_min | float | — | The minimum number of passengers in the bus, i.e., the minimum value of the time series itcs_numberOfPassengers, where NaN values are omitted. |
| itcs_numberOfPassengers_max | float | — | The maximum number of passengers in the bus, i.e., the maximum value of the time series itcs_numberOfPassengers, where NaN values are omitted. |
| status_gridIsAvailable_mean | float | — | The fraction of time the bus is connected to the overhead grid, i.e., the mean value of the time series status_gridIsAvailable. |
| temperature_ambient_mean | float | K | The average ambient temperature, i.e., the mean value of the time series temperature_ambient, where NaN values are omitted. |
| temperature_ambient_min | float | K | The minimum ambient temperature, i.e., the minimum value of the time series temperature_ambient, where NaN values are omitted. |
| temperature_ambient_max | float | K | The maximum ambient temperature, i.e., the maximum value of the time series temperature_ambient, where NaN values are omitted. |

**Table 2.** Column descriptions of the metadata table. The unit column is left blank for non-numeric data types.

- During normal operation, the bus is expected to be at a standstill for some time at both the beginning and the end of each record. If this is not the case, parts of the logging toolchain may have failed to start in time or might have terminated unexpectedly. Thus, we reject 152 missions, where the bus operation does not meet this standstill criterion.
- In some of the records, the bus is found to be not driving at all. This might happen, for instance, if the bus is started during maintenance work. As such records do not represent a regular public transportation operation, they are rejected. This reduces the dataset by 42 missions.
- To exclude any test drives and short missions to, from, or within a depot, we require each driving mission to last at least 3 h. This reduces the dataset by 211 missions.
- During regular operation on the trolley bus routes covered in this dataset, no prolonged standstills are to be expected. Therefore, we filter all missions with any standstill time of more than 30 min. This reduces the dataset by 30 missions.

The selection criteria listed above were established iteratively over the years that we have been working on the data. We believe that these simple criteria are adequate to consistently remove all data records that are contaminated due to software malfunction or that are not representative of a regular public transportation operation, such as drives within a garage or to a workshop. In the following two subsections, we provide some visualizations of the data in the ZTBus dataset[9]. These visualizations offer valuable insights into the dataset's contents and quality. Additionally, they help identify anomalies and outliers, thus guiding the determination of the data selection steps discussed above.

**Time series inspection.** Throughout the activities within our research projects[10,11], we visually analyzed hundreds of time windows that are similar to the one shown in Fig. 4. Such visualizations reveal, for example, the consistency of the wheel speed signals w.r.t. the steering and articulation angles. In particular, a left turn is evident at around second 40, with a negative articulation angle and a positive steering angle, according to the definitions given in Table 1. Accordingly, as expected in a left turn, the right front wheel turns slightly faster than the left.

A slightly coarser time scale is used in Fig. 5, which visualizes the proper synchronization of the VCU signals with the ITCS signals. This figure shows that the times at which the stop names and the passenger counts are reported correspond to the times at which the bus departs from the respective stops.

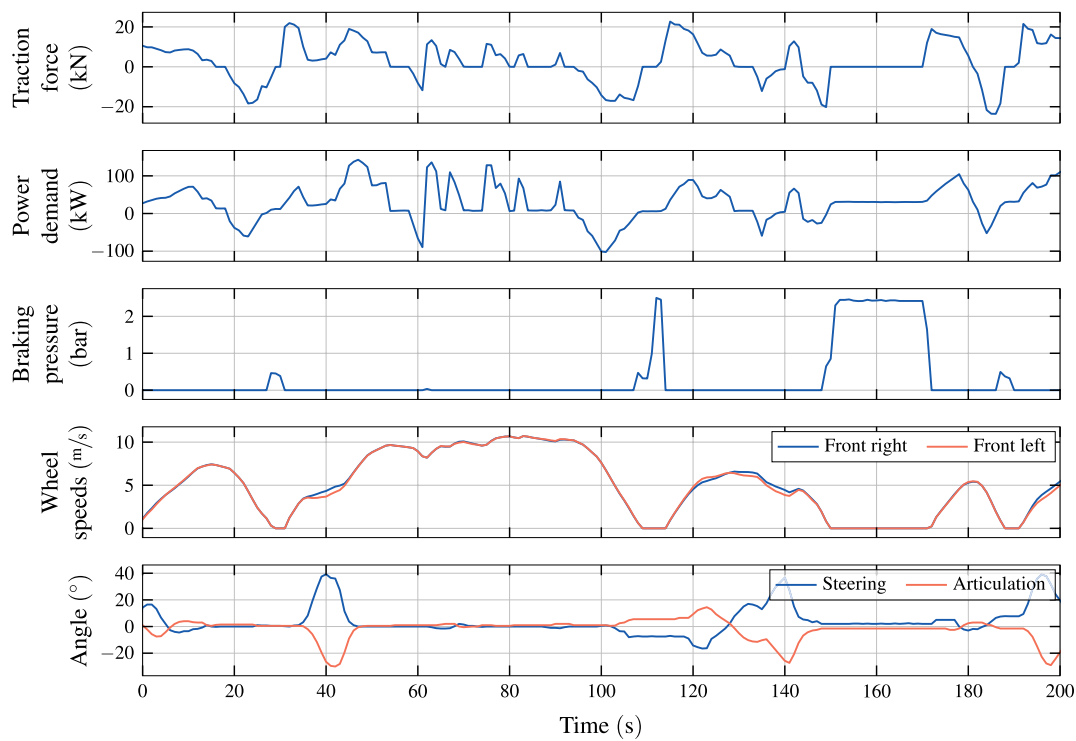

**Fig. 4** Detailed view on example time series recorded in the morning of 7 Mar 2021 on bus numbered 183 on route 72 of VBZ.

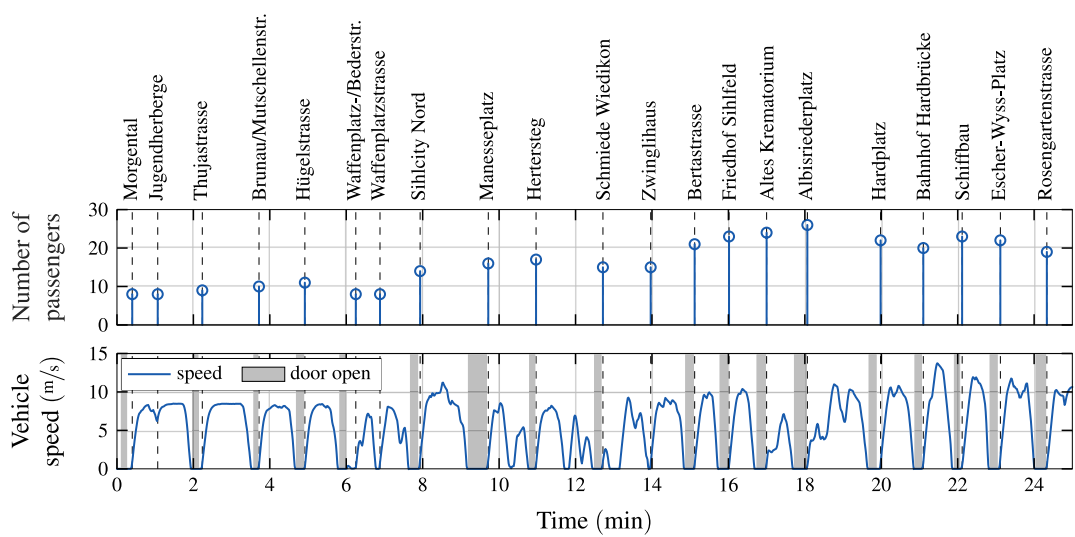

**Fig. 5** Exemplary ITCS data aligned with the speed profile recorded in the morning of 7 Mar 2021 on the bus numbered 183 on route 72 of VBZ. The names of the bus stops are shown above the graph. The areas shaded in gray indicate that at least one door is open.

Some signals are best visualized over the course of the daily operation of a bus, as exemplified in Fig. 6. The repetitive nature of the driving profile is clearly observable in both the pronounced elevation profile of the depicted bus route 72 and the passenger volume. The measured ambient temperature indicates on the one hand that the bus started its mission directly from a depot whose temperature lies significantly above the outside temperature, and on the other hand that the thermal inertia experienced by the sensor is quite significant due to its placement behind bodywork. Due to the sensor resolution, the temperature is only available in increments of 1 K. However, as a result of the linear interpolation used during processing, the dataset may contain single temperature values between these increments.

The GNSS data deliberately was not modified and is provided as raw data, except for the linear interpolation necessary for time synchronization. Therefore, the data might be noisy or imprecise in some locations or may be missing completely during certain time windows. Hence, for certain types of applications, such data may have to be excluded or pre-processed by applying dedicated smoothing or filtering algorithms. Conversely, incomplete and imprecise data can be used as valuable training or validation data, e.g., for dead-reckoning and map-matching algorithms. Anyways, the GNSS data is mostly of good quality, as the exemplary visualization in Fig. 7 clearly shows. Despite the complicated installation of the overhead infrastructure and the footbridges around the crossing depicted, the quality of the raw measurements is by far sufficient to determine which roads were taken.

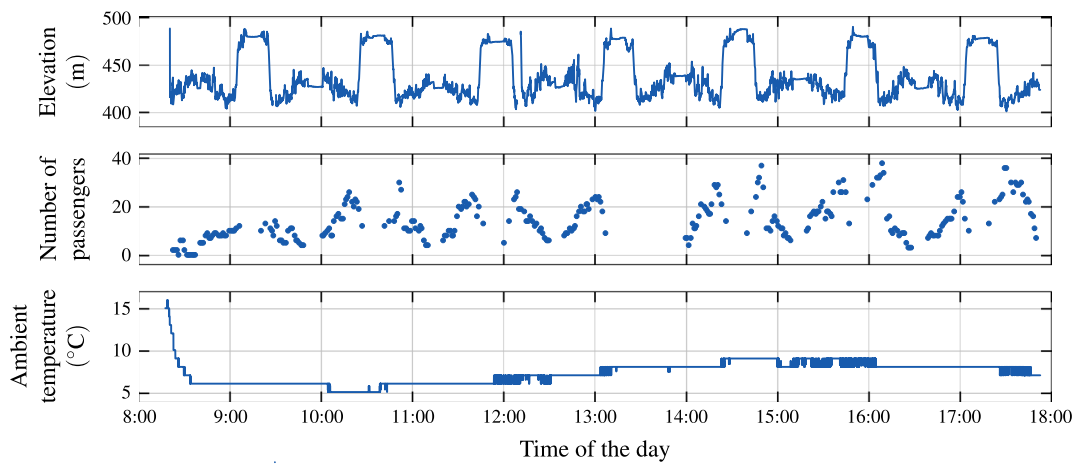

**Fig. 6** Example time series recorded throughout the day on 7 Mar 2021 on the bus numbered 183 on VBZ route 72.

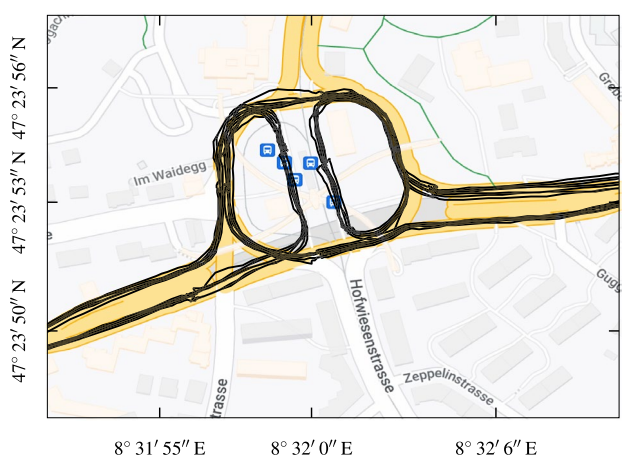

**Fig. 7** Example GNSS trajectories of 7 outbound and return trips each, recorded on 7 Mar 2021 around Bucheggplatz in Zurich, Switzerland. The square is an important transportation hub in the center of Zurich, connecting two tram routes, two bus routes, and three trolley bus routes. Map data ©2023 Google.

**Statistical analysis.** In order verify and validate the integrity of the individual signals, we perform a rudimentary statistical analysis on the large amount of collected data. In particular, we examine a multitude of histograms, three of which exemplified in Fig. 8. Such an analysis of the respective minimum, mean, and maximum values of all driving missions lends itself to detect anomalies and outliers relatively quickly. These visualizations were a helpful tool in the development of the selection criteria mentioned above.

An inspection of the three histograms shown in Fig. 8 reveals that the vehicle speed is slightly negative in some situations. This typically occurs when the vehicle is starting or stopping. From experience, we also know that the average speed of a transit bus in Zurich is around 15 km/h, while the maximum speed rarely exceeds 65 km/h. These facts are well represented and confirmed in the dataset. The distinct peaks shown in the maximum and minimum power demand levels are mainly due to the combined power capacity of the two electric motors of the buses, which is around 320 kW for both negative and positive values. Assuming some auxiliary power consumption in the range of 20 kW to 30 kW, these power limits perfectly explain the peaks observed at around $-300$ kW and 350 kW, respectively. The average power demand is in the range of 15 kW to 35 kW, which corresponds to the expected range of 1.5 kWh/km to 2.0 kWh/km for transit buses driving at the mean vehicle speed mentioned above. Finally, the distribution of the number of passenger shows that the average occupation ranges between 10 and 30 people. On most missions, the vehicle is both empty and about half full at least once.

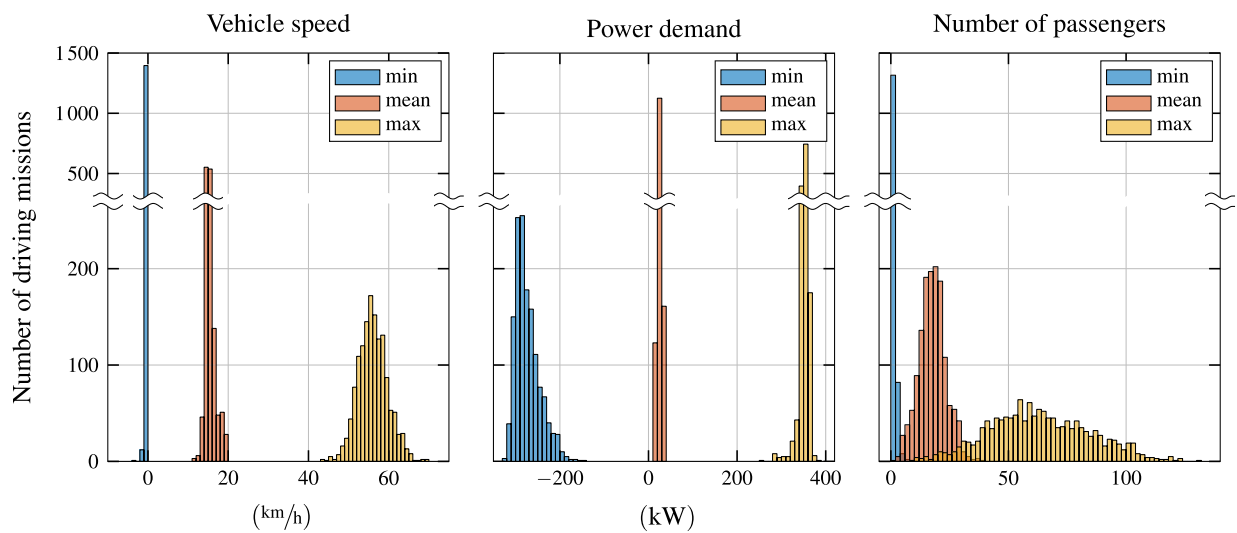

**Fig. 8** Example histograms granting a representative overview of all data values contained in the dataset. Each of the three categories "min", "mean", and "max" refers to the corresponding values of all driving missions, visualized as a histogram. For example, the data values shown for the minimum vehicle speed are the smallest values of `odometry_vehicleSpeed` of each driving mission. To increase the clarity of the graph, the scale of the y-axes changes above the indicated discontinuities.

### Usage Notes
All files of the ZTBus dataset[9] described in this paper are provided in the CSV format using UTF-8 encoding. No special tools are necessary to load or interpret this data and most data processing tools can seamlessly work with data in this format. For convenience, sample Matlab code is provided along with the dataset to recreate the figures shown in this paper. This code can serve as a starting point for numerous new analyses.

### Code availability
The code used to collect, store, filter, and synchronize the data is not published, as it can only be used with the raw data recorded on the specific prototype vehicles, which contains partly proprietary data. As large portions of the code deal with engineering challenges, such as translating data between formats used in different programming languages, ensuring compatibility between software versions, and performing operations in our custom log data base, we do not expect it to be interesting to the readers or useful for any other applications. Instead we directly explain the relevant processing and filtering steps in the respective sections above.

As mentioned above, no specific code is necessary to load or interpret the ZTBus dataset[9]. However, for convenience, the sample Matlab code provided allows to load some parts of the data and recreate most of the figures shown in this manuscript. The code has been developed with Matlab version 9.12 (R2022a) and does not require any specialized toolboxes. It is distributed under the GNU General Public License version 3 (GPLv3) alongside the ZTBus dataset[9].

### Acknowledgements

This work was supported by the SFOE (contract numbers SI/501321-01 and SI/501979-01) and the industrial partners Carrosserie HESS AG and VBZ.

### Author contributions

F.W. and A.R. developed the methodology and software, curated, analyzed, and visualized the data, and wrote the original draft. C.O. was responsible for funding acquisition and supervision. All authors reviewed the manuscript.

### Competing interests

The authors declare no competing interests.

### Additional information

**Correspondence** and requests for materials should be addressed to F.W.

**Reprints and permissions information** is available at www.nature.com/reprints.

**Publisher's note** Springer Nature remains neutral with regard to jurisdictional claims in published maps and institutional affiliations.